\documentclass[reprint,amsmath,amssymb,aps,pra,showpacs]{revtex4-1}

\usepackage{graphicx}
\usepackage{dcolumn}
\usepackage{bm}
\usepackage[utf8]{inputenc}
\usepackage{amsfonts}
\usepackage{tabularx}
\usepackage[normalem]{ulem}

\usepackage{braket} 
\usepackage{nicefrac} 
\usepackage{IEEEtrantools} 
\usepackage{dcolumn}

\newcount\colveccount
\newcommand*\colvec[1]{
        \global\colveccount#1
        \begin{pmatrix}
        \colvecnext
}
\def\colvecnext#1{
        #1
        \global\advance\colveccount-1
        \ifnum\colveccount>0
                \\
                \expandafter\colvecnext
        \else
                \end{pmatrix}
        \fi
}

\begin{document}
\title{Theoretical prediction of the Fine and Hyperfine structure of heavy muonic atoms}

\author{Niklas Michel}\email{nmichel@mpi-hd.mpg.de}
\author{Natalia S. Oreshkina}\email{oresh@mpi-hd.mpg.de}
\author{Christoph H. Keitel}

\affiliation{Max~Planck~Institute for Nuclear Physics, Saupfercheckweg 1,
69117 Heidelberg, Germany}

\date{\today}


\begin{abstract}
Precision calculations of the fine and hyperfine structure of muonic atoms are performed in a relativistic approach and results for muonic $^{205}$Bi, $^{147}$Sm, and $^{89}$Zr are presented. The hyperfine structure due to magnetic dipole and electric quadrupole splitting is calculated in first order perturbation theory, using extended nuclear charge and current distributions. The leading correction from quantum electrodynamics, namely vacuum polarization in Uehling approximation, is included as a potential directly in the Dirac equation. Also, an effective screening potential due to the surrounding electrons is calculated, and the leading relativistic recoil correction is estimated.
\end{abstract}

\maketitle

\section{Introduction}
A muon is a charged elementary particle, which is in many aspects similar to the electron, in particular, it has the same electric charge, but it is ${\approx}\,{200}$ times heavier than the electron~\cite{codata}. When coming close to an atom, a muon can be captured by the nucleus and form a hydrogen-like muonic ion, which is typically also surrounded by the atomic electrons. This atomic system is commonly referred to as a muonic atom. The lifetime of the muon is big enough to be considered stable in the structure calculations of these muonic bound states. Muonic atomic systems feature strong dependence on nuclear parameters and therefore can provide information about atomic nuclei \cite{Wheeler1949}. This triggered interest in precise knowledge of the level structure of muonic atoms \cite{BorieRinker1982,Devons1995}. Due to the muon's high mass, it is located much closer to the nucleus; and, especially for heavy nuclei, this results in big nuclear size effects and a strong dependence of the muon bound state energies on the nuclear charge and current distributions, as well as large relativistic effects.

A combination of the knowledge about the level structure and experiments measuring the transition energies in muonic atoms enabled the determination of nuclear parameters like charge radii \cite{Piller1990,Schaller1980}, quadrupole moments \cite{Dey1979}, and magnetic HFS constants \cite{Ruetschi1984}. One of the most precise measurements to date is the determination of the nuclear root-mean-square radius of $^{208}$Pb on a $0.2\%$ level~\cite{Bergem1988}.

Recent measurements on muonic hydrogen renewed the interest in muonic atoms, revealing a disagreement between the values for the proton charge radius extracted from muonic and electronic systems \cite{Pohl2010}. This allows the assumption that there can be unidentified effects in muonic systems, and triggered detailed theoretical investigation of muonic hydrogen und light muonic atoms, e.g., Refs \cite{indelicato2013,pachucki2015}. Deeper knowledge of the physics of heavy muonic atoms could also contribute to the understanding of the muonic puzzles. In addition, nuclear parameters obtained from muonic x rays would be beneficial for experiments on atomic parity violation \cite{Wansbeek2008}. For this reason, there are upcoming experiments on heavy muonic atoms \cite{kirch2016}. The complicated level structure of these systems demands accurate theoretical calculations.

We present updated state-of-the-art calculations of the fine and hyperfine structure of heavy muonic atoms and analyze the individual contributions. In combination with experimental data, they can be used for the determination and further improvement of values of nuclear parameters. The fine structure is calculated including finite size effects and leading order effects of the vacuum polarization. Additionally, the screening from the surrounding atomic electrons is considered. The hyperfine structure is then calculated with extended quadrupole and magnetization distributions, including the previously mentioned effects. Results are presented for muonic $^{205}$Bi, $^{147}$Sm, and $^{89}$Zr. The dual-kinetic-balance method \cite{Shabaev2004} was applied for the numerical evaluation of the listed contributions.

Muonic relativistic units with ${\hbar}{=}{c}{=}{m_\mu}{=}{1}$ are used, where $m_\mu$ is the muon's mass, and the Heavyside charge unit with $\alpha=e^2/4\pi$, where $\alpha$ is the fine structure constant and the electron's charge is ${e}{<}{0}$.
\section{Interaction between Muon and Nucleus}
The total Hamiltonian for a muon bound to a nucleus can be written as a sum of nuclear, muonic, and interaction Hamiltonian \cite{Devons1995}. Thus, we consider the Hamiltonian
\begin{equation}
H = H_{N} + H^{(0)}_{\mu} + H_{\mu - N},
\end{equation}
with the nuclear Hamiltonian $H_{N}$, the Dirac Hamiltonian $H^{(0)}_{\mu}$ for the free muon, and the interaction Hamiltonian $H_{\mu - N}$. The nucleus is described in the rotational model, i.e. in a state with well defined angular momentum and charge- and current density in the body fixed nuclear frame \cite{Kozhedub2008}. As a next step, the interaction between the bound muon and the atomic nucleus is expanded, where electric and magnetic interactions are taken into account. The interaction Hamiltonian is
\begin{equation}
\label{eq:Hint}
H_{\mu - N} = H_{E} + H_{M}
\end{equation}
where the electric part reads
\begin{equation}
\label{eq:elInt}
H_{E}= - \alpha \int \mathrm{d}V^{\prime}\, \frac{\rho (\vec{r}^{\,\prime})}{|\vec{r}_{\mu}-\vec{r}^{\,\prime}|} ,
\end{equation}
with the fine structure constant $\alpha$, the position $\vec{r}^{\,\prime}$ of the nuclear charge distribution and the position $\vec{r}_{\mu}^{\,\prime}$ of the muon in the nuclear frame. The nuclear charge distribution $\rho(\vec{r})$ is normalized to the nuclear charge $Z$ as
\begin{equation}
\label{eq:norm}
\int \mathrm{d}V\rho(\vec{r}) = Z.
\end{equation}
Conveniently, the nuclear charge distribution is divided into a spherically symmetric part $\rho_0(r)$ and a part $\rho_2(r)$ describing the quadrupole distribution in the nuclear frame as \cite{hitlin1970}
\begin{equation}
\label{eq:rho}
\rho(\vec{r}^{\,\prime}) = \rho_0(r^{\prime}) + \rho_2(r^{\prime}) \, Y_{20}(\vartheta^\prime,\varphi^\prime),
\end{equation}
with the spherical harmonics $Y_{lm}(\vartheta,\varphi)$. Since an analogous part for the dipole distribution would be an operator of odd parity, it would vanish after averaging with muon wave functions of defined parity \cite{johnson2007}, and thus it is not considered here and neither are higher multipoles beyond the quadrupole term. Correspondingly, the electric interaction Hamiltonian from (\ref{eq:Hint}) can be written as
\begin{equation}
\label{eq:quadInt}
H_E = H^{(0)}_E + H^{(2)}_E,
\end{equation}
where the spherically symmetric part of the charge distribution gives rise to
\begin{equation}
\label{eq:Hmonopole}
H^{(0)}_E(r_\mu)= - 4 \pi \alpha \int_0^\infty \mathrm{d}r \, r^2 \frac{\rho_0(r)}{r_>},
\end{equation}
with $r_>=\text{max}(r,r_\mu)$. This interaction Hamiltonian will be included in the numerical solution of the Dirac equation for the muon as described in Sec. \ref{sec:radialEq}. The quadrupole part of the interaction $H^{(2)}_E$ causes hyperfine splitting, which is calculated perturbatively in Sec. \ref{sec:elQuad}.\\

As for the magnetic part, we consider dipole interaction. Therefore, the corresponding interaction Hamiltonian from (\ref{eq:Hint}) reads \cite{Elizarov2005}
\begin{equation}
\label{eq:Hmag}
H_{M} = \frac{|e|}{4 \pi}\,\vec{\mu}\cdot \left( F_{\text{BW}}(r) \frac{\vec{r}}{r^3} \times \vec{\alpha} \right),
\end{equation}
with the charge of the muon $e=-|e|$, the nuclear magnetic moment $\vec{\mu}$, its distribution function $F_{\text{BW}}$, and the Dirac matrices $\vec{\alpha}$. If the nuclear current density is described by a normalized scalar function $f_\mu(r)$ as
\begin{equation}
\label{eq:currentdistr}
\vec{j}(r)= \text{rot}\left(\vec{\mu}f_\mu(r)\right),
\end{equation}
then the distribution function is given by
\begin{equation}
\label{eq:Fbw}
F_{\text{BW}}(r)=-r^2 \frac{\partial}{\partial r}\,\int \text{d}V^{\prime}\,\frac{f_\mu(r^{\prime})}{|\vec{r}-\vec{r}\,^{\prime}|}.
\end{equation}
The difference in the hyperfine splitting between a point-like magnetic moment and a spacial distribution of the magnetization is called the Bohr-Weisskopf effect \cite{bohrWeisskopf1950}. In Sec. \ref{sec:elQuad}, the matrix elements of the magnetic interaction are analyzed, paying special attention to the distribution function $F_{\text{BW}}$. We expect the contribution of the higher-order terms, namely electric octupole, magnetic quadrupole, and beyond, to be smaller than the uncertainty of the considered terms \cite{Devons1995,Steffen1985}. Therefore they can be ignored here.

For evaluating these Hamiltonians, the appropriate states are states of defined total angular momentum. A nuclear state $\ket{IM}$ with nuclear angular momentum quantum number $I$ and projection $M$ on the $z$ axis of the laboratory frame and a muonic state $\ket{n\kappa m}$ with total angular momentum $j(\kappa)=|\kappa|-\frac{1}{2}$ and projection $m$ are coupled to a state $\ket{FM_FI\kappa}$ with angular momentum $F$ and projection $M_F$ as
\begin{equation}
\label{eq:totalState}
\ket{FM_FI\kappa}=\sum_{M,m} C^{FM_F}_{IM\,jm} \ket{I M} \, \ket{n\kappa m},
\end{equation}
where $C^{jm}_{j_1m_1j_2m_2}$ are the Clebsch-Gordan coefficients \cite{varshalovich1988}. Here, $n$ is the principal quantum number of the muon and $\kappa=(-1)^{j+l+\frac{1}{2}}(j+\frac{1}{2})$ with the orbital angular momentum quantum number $l$.
\section{Dirac equation with finite size corrections}
\label{sec:radialEq}
As a basis for further calculations, the Dirac equation
\begin{equation}
\label{eq:diracSph}
\left( \vec{\alpha}\cdot \vec{p}+ \beta + V(r_\mu) \right) \ket{n \kappa m} = (1-E_{n \kappa}) \ket{n \kappa m}
\end{equation} 
is solved for the muon. Here, $\vec{\alpha}$ and $\beta$ are the four Dirac matrices, $E_{n \kappa}$ are the binding energies, and the potential $V(r)$ is the spherically symmetric part of the interaction with the nucleus, which is the monopole contribution from the electric interaction (\ref{eq:Hmonopole}) and the Uehling potential from (\ref{eq:uehl}). A Fermi type charge distribution \cite{Beier2000} is used to model the monopole charge distribution as
\begin{equation}
\label{eq:fermi}
\rho_0 (r)=\frac{N}{1+\text{exp}((r-c)/a)},
\end{equation}
where $a$ is a skin thickness parameter and $c$ the half-density radius. The normalization constant $N$ is chosen such that (\ref{eq:norm}) is fulfilled. It has been proven, that $a=t/(4\,\text{log}3)$, with $t=2.30\,\text{fm}$, is a good approximation for most of the nuclei \cite{Beier2000}. The parameter $c$ is then determined by demanding, that the charge radius squared
\begin{equation}
\left<r^2\right>=\cfrac{\int\text{d}r \, r^4\rho_0(r)}{\int\text{d}r \, r^2\rho_0(r)}
\end{equation}
agrees with the values from the literature \cite{Angeli2013}. Since the potential in (\ref{eq:diracSph}) is spherically symmetric, the angular part can be separated and the solution with spherical spinors $\Omega_{\kappa m}(\vartheta,\varphi)$ can be written as \cite{greiner2000}
\begin{equation}
\ket{n\kappa m}=\frac{1}{r}\colvec{2}{G_{n\kappa}(r)\,\Omega_{\kappa m}}{i\,F_{n\kappa}(r)\,\Omega_{-\kappa m}},
\end{equation}
and the resulting equations for the radial functions are solved with the dual-kinetic-balance method \cite{Shabaev2004} to obtain $G_{n\kappa}$ and $F_{n\kappa}$, and the corresponding eigenenergies numerically. 

In Table \ref{tab:sphDirac}, the binding energies for muonic $^{205}_{83}$Bi, $^{147}_{62}$Sm, and $^{89}_{40}$Zr are shown, both with and without the corrections from the Uehling potential (\ref{eq:uehl}). The finite nuclear size effect is illustrated by also including the binding energies $E^{(C)}_{n\kappa}$ of the pure Coulomb potential $-Z\alpha / r_\mu$, which read \cite{greiner2000}
\begin{equation}
\label{eq:finestructure_formula}
E^{(C)}_{n\kappa}=1-\left(1+\frac{(Z\alpha)^2}{\left( n-|\kappa|+\sqrt{\kappa^2-(Z\alpha)^2} \right)^2}\right)^{-\frac{1}{2}}.
\end{equation}
The uncertainties include the error in the rms radius value as well as a model error, which is estimated via the difference of the binding energies with the Fermi potential (\ref{eq:fermi}) and the potential of a charged sphere with the same rms radius. For heavy nuclei, the finite nuclear size correction can amount up to 50$\,\%$, and thus the binding energy is halved.

\begin{table}[b]
\caption{\label{tab:sphDirac}
Overview of the binding energies for muonic $^{205}_{83}$Bi, $^{147}_{62}$Sm, and $^{89}_{40}$Zr, obtained by solving the Dirac equation with the spherically symmetric parts of the muon-nucleus interaction. The values for solving the Dirac equation only with the electric monopole potential, and with the electric monopole potential and the Uehling potential are presented to show the influence of the leading order vacuum polarization. The binding energies (\ref{eq:finestructure_formula}) for a  point like nucleus are shown as well. The reduced mass is used to include the non-relativistic recoil corrections from Section \ref{sec:recoil}. The corrections from section \ref{sec:screen} are not included in this table. All energies are in keV.}
\begin{ruledtabular}
\begin{tabular}{cclll}
& state & \text{point like}& \text{finite size (fs)}\footnotemark[1] &\text{fs+Uehling}\footnotemark[2]\\ \hline \\[-7pt]
$^{205}$Bi & 1s\nicefrac{1}{2} &\text{21573.3} & \text{10699.(51.)} &\text{10767.(52.)} \\
  & 2s\nicefrac{1}{2} & \text{\phantom{1}5538.6} & \text{\phantom{1}3654.(15.)} & \text{\phantom{1}3674.(15.)}\\
  & 2p\nicefrac{1}{2} & \text{\phantom{1}5538.6} & \text{\phantom{1}4893.(3.)} & \text{\phantom{1}4927.(3.)} \\
  & 2p\nicefrac{3}{2} & \text{\phantom{1}4958.9} & \text{\phantom{1}4706.(5.)} & \text{\phantom{1}4737.(5.)} \\
  & 3s\nicefrac{1}{2} & \text{\phantom{1}2394.3} & \text{\phantom{1}1796.(5.)} & \text{\phantom{1}1804.(6.)} \\
  & 3p\nicefrac{1}{2} & \text{\phantom{1}2394.3} & \text{\phantom{1}2170.0(5)} & \text{\phantom{1}2190.1(5)} \\
  & 3p\nicefrac{3}{2} & \text{\phantom{1}2221.4} & \text{\phantom{1}2131.(1.)} & \text{\phantom{1}2141.(1.)} \\
  & 3d\nicefrac{3}{2} & \text{\phantom{1}2221.4} & \text{\phantom{1}2216.9(3)}& \text{\phantom{1}2227.8(3)}\\
  & 3d\nicefrac{5}{2} & \text{\phantom{1}2174.6} & \text{\phantom{1}2172.8(2)} & \text{\phantom{1}2183.0(2)} \\[7pt]
 $^{147}$Sm & 1s\nicefrac{1}{2} & \text{11423.8} & \text{\phantom{1}7165.(28.)} & \text{\phantom{1}7213.(29.)} \\
  & 2s\nicefrac{1}{2} & \text{\phantom{1}2895.7} & \text{\phantom{1}2230.(7.)} & \text{\phantom{1}2242.(7.)} \\
  & 2p\nicefrac{1}{2} & \text{\phantom{1}2895.7} & \text{\phantom{1}2778.(2.)} & \text{\phantom{1}2795.(2.)} \\
  & 2p\nicefrac{3}{2} & \text{\phantom{1}2736.9} & \text{\phantom{1}2689.(2.)} & \text{\phantom{1}2706.(2.)} \\
  & 3s\nicefrac{1}{2} & \text{\phantom{1}1268.9} & \text{\phantom{1}1061.(2.)} & \text{\phantom{1}1066.(2.)} \\
  & 3p\nicefrac{1}{2} & \text{\phantom{1}1268.9} & \text{\phantom{1}1228.6(4)} & \text{\phantom{1}1234.2(4)} \\
  & 3p\nicefrac{3}{2} & \text{\phantom{1}1221.7} & \text{\phantom{1}1204.7(6)} & \text{\phantom{1}1210.0(6)} \\
  & 3d\nicefrac{3}{2} & \text{\phantom{1}1221.7} & \text{\phantom{1}1221.4(1)} & \text{\phantom{1}1226.2(1)} \\
  & 3d\nicefrac{5}{2} & \text{\phantom{1}1207.6} & \text{\phantom{1}1207.4} & \text{\phantom{1}1212.1} \\[7pt]
 $^{89}$Zr & 1s\nicefrac{1}{2} & \text{\phantom{1}4595.5} & \text{\phantom{1}3643.(8.)} & \text{\phantom{1}3669.(8.)} \\
  & 2s\nicefrac{1}{2} & \text{\phantom{1}1155.2} & \text{\phantom{1}1021.(2.)} & \text{\phantom{1}1026.(2.)} \\
  & 2p\nicefrac{1}{2} & \text{\phantom{1}1155.2} & \text{\phantom{1}1147.8(2)} & \text{\phantom{1}1153.7(2)} \\
  & 2p\nicefrac{3}{2} & \text{\phantom{1}1129.9} & \text{\phantom{1}1127.0(2)} & \text{\phantom{1}1132.6(2)} \\
  & 3s\nicefrac{1}{2} & \text{\phantom{11}510.6} & \text{\phantom{11}469.8(5)} & \text{\phantom{11}471.4(5)} \\
  & 3p\nicefrac{1}{2} & \text{\phantom{11}510.6} & \text{\phantom{11}508.0(1)} & \text{\phantom{11}509.8(1)} \\
  & 3p\nicefrac{3}{2} & \text{\phantom{11}503.1} & \text{\phantom{11}502.0(1)} & \text{\phantom{11}503.8(1)} \\
  & 3d\nicefrac{3}{2} & \text{\phantom{11}503.1} & \text{\phantom{11}503.1} & \text{\phantom{11}504.5} \\
  & 3d\nicefrac{5}{2} & \text{\phantom{11}500.7} & \text{\phantom{11}500.7} & \text{\phantom{11}502.1} \\

\end{tabular}
\end{ruledtabular}
\footnotetext[1]{$V(r_\mu)=H^{(0)}_E(r_\mu)$}
\footnotetext[2]{$V(r_\mu)=H^{(0)}_E(r_\mu)+V_{\text{Uehl}}(r_\mu)$
\\see eq. (\ref{eq:Hmonopole}), (\ref{eq:diracSph}), and (\ref{eq:uehl}) for definitions}
\end{table}
%
%
\section{Vacuum polarization}
\label{sec:qed}

For atomic electrons, usually the self-energy QED correction is comparable to the vacuum polarization correction \cite{Beier2000}. For muons, however, the vacuum polarization correction is much larger due to virtual electron-positron pairs, which are less suppressed due to their low mass compared to the muon's mass \cite{BorieRinker1982}. The spherically symmetric part of the vacuum polarization to first order in $\alpha$ and $Z\alpha$ is the Uehling potential \cite{Elizarov2005}
\begin{align}
V_{\text{Uehl}}(r_\mu)=-\alpha \frac{2\alpha}{3\pi}\int_0^\infty \text{d}r^{\prime}\,4\pi \rho_0(r^\prime)\int_1^\infty \text{d}t\,\left( 1+\frac{1}{2t^2} \right)\nonumber\\
\times\frac{\sqrt{t^2-1}}{t^2} \frac{\text{exp}(-2m_e|r_\mu-r^\prime|t)-\text{exp}(-2m_e(r_\mu+r^\prime)t)}{4m_er_\mu t},
\label{eq:uehl}
\end{align}
where $m_e$ is the electron mass and $\rho_0$ is the spherically symmetric part of the charge distribution from (\ref{eq:rho}). This potential can be directly added to the Dirac equation (\ref{eq:diracSph}). In this way, all iterations of the Uehling potential are included \cite{indelicato2013}. Results for our calculations can be found in Table \ref{tab:sphDirac}.
\section{Recoil corrections}
\label{sec:recoil}
Taking into account the finite mass and the resulting motion of the nucleus leads to recoil corrections to the bound muon energy levels. In nonrelativistic quantum mechanics, as in classical mechanics, the problem of describing two interacting particles can be reduced to a one particle problem by using the reduced mass $m_r$ of the muon-nucleus system \cite{landaulifshitz3}. With the mass of the nucleus $m_N$, the reduced mass reads in the chosen system of units as
\begin{equation}
\label{eq:redmass}
m_r=\frac{m_N}{m_N+1},
\end{equation}
and the Dirac equation is accordingly modified to
\begin{equation}
\label{eq:diracSphRed}
\left( \vec{\alpha}\cdot \vec{p}+ \beta\,m_r + V(r_\mu) \right) \ket{n \kappa m} = (m_r-E_{n \kappa}) \ket{n \kappa m}.
\end{equation} 
In relativistic quantum mechanics, this separation is not possible. We follow an approach used in Refs. \cite{friar1973,BorieRinker1982}, which includes the nonrelativistic part of the recoil correction already in the wave functions by using the reduced mass in the Dirac equation and calculating the leading relativistic corrections perturbatively. If $E^{\text{(fm)}}_{n\kappa}$ denotes the binding energy of (\ref{eq:diracSph}) with the finite size potential (\ref{eq:Hmonopole}) but with the reduced mass replaced by the full muon rest mass, and $E^{\text{(rm)}}_{n\kappa}$ the binding energy in the same potential but with the reduced mass (\ref{eq:redmass}), then the leading relativistic recoil correction $\Delta E^{\text{(rec,rel)}}_{n\kappa}$ according to Ref. \cite{BorieRinker1982} reads
\begin{equation}
\label{eq:relrec}
\Delta E^{\text{(rec,rel)}}_{n\kappa} = -\frac{\left(E^{\text{(fm)}}_{n\kappa}\right)^2}{2 M_N}+\frac{1}{2 M_N}\left< h(r) + 2 E^{\text{(fm)}}_{n\kappa} P_1(r)  \right>,
\end{equation}
where $M_N$ is the mass of the nucleus, and the functions $h(r)$ and $P_1(r)$ are defined in Eqs. (109) and (111) of Ref. \cite{BorieRinker1982}, respectively. In Table \ref{tab:recoil}, the binding energies obtained from solving the Dirac equation with the muon rest mass and the reduced mass of the muon-nucleus system are compared, and the leading relativistic recoil correction is shown. The uncertainties include errors in the rms radius, the model of the charge distribution and for the relativistic recoil, and a $(m_\mu/M_N)^2$ term due to higher-order corrections in the mass ratio of muon and nucleus, which dominates the uncertainty for lower $Z$.
\begin{table}
\caption{\label{tab:recoil}Recoil corrections to the binding energies of the muon. fm (full mass) denotes the finite size binding energy, analogous to the fourth column of Table \ref{tab:sphDirac}, but with the rest mass of the muon used in the Dirac equation. $\Delta E_{\text{rec,nr}}$ is the non-relativistic recoil correction, which is the difference between the finite size Dirac solutions with reduced mass and full mass, respectively. $\Delta E^{\text{(rec,rel)}}_{n\kappa}$ is the leading relativistic recoil correction from Section \ref{sec:recoil}.
All energies are in keV.}
\begin{ruledtabular}
\begin{tabular}{lllll}
& state & $E^{\text{(fm)}}$ &$\Delta E^{\text{rec,nr}}$&$\Delta E^{\text{(rec,rel)}}_{n\kappa}$\footnotemark[1]\\ \hline \\[-7pt]
 $^{205}$Bi & 1s\nicefrac{1}{2} & \text{10702.(51.)} & \text{-2.80(4)} & \text{0.39(4)} \\
  & 2s\nicefrac{1}{2} & \text{\phantom{1}3656.(15.)} & \text{-1.42(2)} & \text{0.09(3)}\\
  & 2p\nicefrac{1}{2} & \text{\phantom{1}4895.6(3.0)} & \text{-2.24(1)} & \text{0.12(3)} \\
  & 2p\nicefrac{3}{2} & \text{\phantom{1}4708.2(4.6)} & \text{-2.27(1)} & \text{0.01(1)} \\
  & 3s\nicefrac{1}{2} & \text{\phantom{1}1796.6(5.5)} & \text{-0.78(1)} & \text{0.03(3)} \\
  & 3p\nicefrac{1}{2} & \text{\phantom{1}2180.0(0.5)} & \text{-1.05} & \text{0.03(3)} \\
  & 3p\nicefrac{3}{2} & \text{\phantom{1}2131.9(1.3)} & \text{-1.06} & \text{0.03(3)} \\
  & 3d\nicefrac{3}{2} & \text{\phantom{1}2218.1(0.3)} & \text{-1.21} & \text{0.02(2)} \\
  & 3d\nicefrac{5}{2} & \text{\phantom{1}2174.0(0.2)} & \text{-1.19} & \text{0.02(2)} \\[7pt]
 $^{147}$Sm & 1s\nicefrac{1}{2} & \text{\phantom{1}7168.(28.)} & \text{-3.17(4)} & \text{0.29(7)} \\
  & 2s\nicefrac{1}{2} & \text{\phantom{1}2231.1(6.7)} & \text{-1.31(1)} & \text{0.05(5)} \\
  & 2p\nicefrac{1}{2} & \text{\phantom{1}2779.4(1.5)} & \text{-1.97(1)} & \text{0.05(5)} \\
  & 2p\nicefrac{3}{2} & \text{\phantom{1}2691.2(1.8)} & \text{-1.96(1)} & \text{0.04(4)} \\
  & 3s\nicefrac{1}{2} & \text{\phantom{1}1062.0(2.3)} & \text{-0.68(1)} & \text{0.02(2)} \\
  & 3p\nicefrac{1}{2} & \text{\phantom{1}1229.5(0.4)} & \text{-0.89} & \text{0.01(1)} \\
  & 3p\nicefrac{3}{2} & \text{\phantom{1}1205.6(0.6)} & \text{-0.89} & \text{0.01(1)} \\
  & 3d\nicefrac{3}{2} & \text{\phantom{1}1222.3(0.1)} & \text{-0.93} & \text{0.01(1)} \\
  & 3d\nicefrac{5}{2} & \text{\phantom{1}1208.3} & \text{-0.92} & \text{0.01(1)} \\[7pt]
 $^{89}$Zr & 1s\nicefrac{1}{2} & \text{\phantom{1}3646.5(8.2)} & \text{-3.36(3)} & \text{0.15(15)} \\
  & 2s\nicefrac{1}{2} & \text{\phantom{1}1022.4(1.5)} & \text{-1.11(1)} & \text{0.02(2)} \\
  & 2p\nicefrac{1}{2} & \text{\phantom{1}1149.2(0.2)} & \text{-1.43} & \text{0.01(1)} \\
  & 2p\nicefrac{3}{2} & \text{\phantom{1}1128.4(0.2)} & \text{-1.41} & \text{0.01(1)} \\
  & 3s\nicefrac{1}{2} & \text{\phantom{11}470.3(0.5)} & \text{-0.54} & \text{0.01(1)} \\
  & 3p\nicefrac{1}{2} & \text{\phantom{11}508.6(0.1)} & \text{-0.64} & \text{0.00} \\
  & 3p\nicefrac{3}{2} & \text{\phantom{11}502.7(0.1)} & \text{-0.63} & \text{0.00} \\
  & 3d\nicefrac{3}{2} & \text{\phantom{11}503.7} & \text{-0.64} & \text{0.00} \\
  & 3d\nicefrac{5}{2} & \text{\phantom{11}501.3} & \text{-0.63} & \text{0.00} \\
\end{tabular}
\end{ruledtabular}
\footnotetext[1]{$\Delta E^{\text{rec,nr}}:=E^{\text{(red.mass)}}-E^{\text{(fm)}}$, see Section \ref{sec:recoil} for definitions.}
\end{table}
%
%
\section{Electron screening}
\label{sec:screen}
\begin{table}
\caption{\label{tab:screen}Electron screening corrections to the bound muon energy levels. $\Delta E_{\rm{S,eff}}^{(1)}$ and $\Delta E_{\rm{S,eff}}^{(1+2)}$ are the screening corrections with the effective nuclear charge method, whereas $\Delta E_{\rm{S,3step}}^{(1)}$ and $\Delta E_{\rm{S,3step}}^{(1+2)}$ use the 3 step calculation, both described in Section \ref{sec:screen}. For the superscript $(1)$, only the 1s electrons are considered, while for $({1}{+}{2})$, all electrons from the first and second shell are considered. All energies are in keV.}
\begin{ruledtabular}
\begin{tabular}{ccrrrr}
&$\mu$-state & $\Delta E_{\rm{S,eff}}^{(1)}$  & $\Delta E_{\rm{S,eff}}^{(1+2)}$ & $\Delta E_{\rm{S,3step}}^{(1)}$ & $\Delta E_{\rm{S,3step}}^{(1+2)}$\\ \hline \\[-7pt]
 $^{205}$Bi & 1s\nicefrac{1}{2} & \text{5.555} & \text{10.825} & \text{5.555} & \text{10.825} \\
  & 2s\nicefrac{1}{2} & \text{5.537} & \text{10.803} & \text{5.538} & \text{10.805} \\
  & 2p\nicefrac{1}{2} & \text{5.548} & \text{10.817} & \text{5.549} & \text{10.818} \\
  & 2p\nicefrac{3}{2} & \text{5.547} & \text{10.816} & \text{5.548} & \text{10.817} \\
  & 3s\nicefrac{1}{2} & \text{5.490} & \text{10.748} & \text{5.494} & \text{10.753} \\
  & 3p\nicefrac{1}{2} & \text{5.514} & \text{10.776} & \text{5.516} & \text{10.779} \\
  & 3p\nicefrac{3}{2} & \text{5.512} & \text{10.774} & \text{5.515} & \text{10.777} \\
  & 3d\nicefrac{3}{2} & \text{5.526} & \text{10.791} & \text{5.528} & \text{10.793} \\
  & 3d\nicefrac{5}{2} & \text{5.525} & \text{10.789} & \text{5.527} & \text{10.792} \\[7pt]
 $^{147}$Sm & 1s\nicefrac{1}{2} & \text{3.705} & \text{7.312} & \text{3.705} & \text{7.312} \\
  & 2s\nicefrac{1}{2} & \text{3.699} & \text{7.305} & \text{3.700} & \text{7.305} \\
  & 2p\nicefrac{1}{2} & \text{3.703} & \text{7.309} & \text{3.703} & \text{7.309} \\
  & 2p\nicefrac{3}{2} & \text{3.703} & \text{7.309} & \text{3.703} & \text{7.309} \\
  & 3s\nicefrac{1}{2} & \text{3.682} & \text{7.285} & \text{3.683} & \text{7.286} \\
  & 3p\nicefrac{1}{2} & \text{3.689} & \text{7.293} & \text{3.691} & \text{7.295} \\
  & 3p\nicefrac{3}{2} & \text{3.689} & \text{7.293} & \text{3.690} & \text{7.294} \\
  & 3d\nicefrac{3}{2} & \text{3.694} & \text{7.299} & \text{3.695} & \text{7.300} \\
  & 3d\nicefrac{5}{2} & \text{3.694} & \text{7.298} & \text{3.694} & \text{7.299} \\[7pt]
 $^{89}$Zr & 1s\nicefrac{1}{2} & \text{2.214} & \text{4.405} & \text{2.214} & \text{4.405} \\
  & 2s\nicefrac{1}{2} & \text{2.212} & \text{4.402} & \text{2.212} & \text{4.403} \\
  & 2p\nicefrac{1}{2} & \text{2.213} & \text{4.403} & \text{2.213} & \text{4.403} \\
  & 2p\nicefrac{3}{2} & \text{2.213} & \text{4.403} & \text{2.213} & \text{4.403} \\
  & 3s\nicefrac{1}{2} & \text{2.205} & \text{4.395} & \text{2.206} & \text{4.396} \\
  & 3p\nicefrac{1}{2} & \text{2.207} & \text{4.397} & \text{2.208} & \text{4.398} \\
  & 3p\nicefrac{3}{2} & \text{2.207} & \text{4.397} & \text{2.208} & \text{4.398} \\
  & 3d\nicefrac{3}{2} & \text{2.209} & \text{4.399} & \text{2.210} & \text{4.400} \\
  & 3d\nicefrac{5}{2} & \text{2.209} & \text{4.399} & \text{2.209} & \text{4.400} \\

\end{tabular}
\end{ruledtabular}
\end{table}
The effect of the surrounding electrons on the binding energies of the muon was estimated following Ref. \cite{vogel1973} by calculating an effective screening potential from the charge distribution of the electrons as
\begin{equation}
\label{eq:screenPot}
V_{e}(\vec{r}_\mu)=-\alpha \int \mathrm{d}V\frac{\rho_e (\vec{r})}{|\vec{r}_\mu-\vec{r}|},
\end{equation}
and using this potential in the Dirac equation for the muon. The charge distribution of the electrons is obtained by their Dirac wave functions as $\rho_e (\vec{r})=\sum_i \psi_{e_i}^*(\vec{r})\cdot \psi_{e_i}(\vec{r})$, where $\psi_{e_i}(\vec{r})$  is the four component spinor of the $i$-th considered electron. In order to obtain the wave functions of the electrons, it has to be taken into account, that the muon essentially screens one unit of charge from the nucleus. The simplest possibility is to replace the nuclear charge by an effective charge $\tilde{Z}=Z-1$ and then solve the Dirac equation for the electron with this modified nuclear potential. Another possibility is to start solving the Dirac equation for the muon in the nuclear potential without electron screening. Then, the Dirac equation for the electron is solved for all required states, adding the screening potential due to the bound muon
\begin{equation}
V_{\mu}(\vec{r}_e)=-\alpha \int \mathrm{d}V\frac{\psi_\mu^*(\vec{r})\cdot \psi_\mu(\vec{r})}{|\vec{r}_e-\vec{r}|},
\end{equation}
analogously to (\ref{eq:screenPot}).
The interaction between the electrons is not taken into account here. Finally, the Dirac equation for the muon is solved again, now including the nuclear potential and the screening potential (\ref{eq:screenPot}) due the atomic electrons from the considered electron configuration. This procedure can be repeated in the spirit of Hartree's method \cite{bethe_salpeter} until the electrons and the muon are self-consistent in the fields of each other, but our studies show that one iteration is usually enough since the overlap of muon and electron wave functions in heavy muonic atoms is small. It is important to note, that here the screening potential depends to a small extent on the state of the muon, since the muon wave function is used in the calculation for the electron wave function. The atomic electrons primarily behave like a charged shell around the muon and the nucleus; thus every muon level is mainly shifted by a constant term, which is not observable in muonic transitions. The screening correction $\Delta E_S$ is defined as the difference of the binding energy without screening potential and with screening potential, therefore a positive value indicates that the muon is less bound due to the screening effect. The main contribution to the nonconstant part of the screening potential comes from the 1$s$ electrons, since their wave functions have the biggest overlap with the muon; therefore the exact electron configuration has only a minor effect on transition energies \cite{vogel1973}. In Table \ref{tab:screen}, results for the screening correction are shown for both mentioned methods and for different electron configurations. Values of the screening correction for different electron configurations show that a 10\% error for the non-constant part is a reasonable estimate.
\section{Hyperfine interactions}
\label{sec:hfs}
\subsection{Electric quadrupole splitting}
\label{sec:elQuad}
\begin{table*}
\caption{\label{tab:hfs}
Results for the electric quadrupole and magnetic dipole hyperfine splitting for a selection of hyperfine states of muonic $^{205}_{83}$Bi ($I=\frac{9}{2}$), $^{147}_{62}$Sm ($I=\frac{7}{2}$), and $^{89}_{40}$Zr ($I=\frac{9}{2}$). $\braket{H_E^{(2)}}$ are the values of the electric quadrupole splitting. $\braket{H_M^{\rm{hom}}}$ is the magnetic dipole splitting from (\ref{eq:hmag}) using a homogeneous nuclear current distribution and $\braket{H_M^{\rm{sp}}}$ using the nuclear magnetization distribution in the single particle model. See Section \ref{sec:hfs} for definitions. All energies are in keV.}
\begin{ruledtabular}
\begin{tabular}{ccllllll}
 nucleus&state&\multicolumn{2}{c}{$\braket{H_E^{(2)}}$}&\multicolumn{2}{c}{$\braket{H_M^{\rm{hom}}}$}&\multicolumn{2}{c}{$\braket{H_M^{\rm{sp}}}$}\\
 & &$F=I-\frac{1}{2}$&$F=I+\frac{1}{2}$&$F=I-\frac{1}{2}$&$F=I+\frac{1}{2}$&$F=I-\frac{1}{2}$&$F=I+\frac{1}{2}$\\[2pt] \hline \\[-7pt]
   $^{205}$Bi & 1s\nicefrac{1}{2} & \phantom{-11}0 & \phantom{-11}0 & -2.27(20) &\phantom{-}1.86(16) & -2.41(20) &\phantom{-}1.97(16) \\
  & 2s\nicefrac{1}{2} & \phantom{-11}0 & \phantom{-11}0 & \text{-0.43(5)} &\phantom{-}0.35(4) & -0.47(6) &\phantom{-}0.38(4) \\
  & 2p\nicefrac{1}{2} & \phantom{-11}0 & \phantom{-11}0 & -1.23(11) & \phantom{-}1.01(9) & -1.31(11) &\phantom{-}1.07(10) \\
  & 2p\nicefrac{3}{2} & -175.(24.) & \phantom{-}175.(24.) & -0.55(2) & \phantom{-}0.010(4) & -0.554(22) & \phantom{-}0.098(4) \\
  & 3s\nicefrac{1}{2} & \phantom{-11}0 & \phantom{-11}0 & \text{-0.144(20)} & \phantom{-}0.118(16) & -0.160(20) & \phantom{-}0.131(16) \\
  & 3p\nicefrac{1}{2} & \phantom{-11}0 & \phantom{-11}0 & -0.311(33) & \phantom{-}0.255(26) & -0.336(33) & \phantom{-}0.275(27) \\
  & 3p\nicefrac{3}{2} & \phantom{1}-48.9(8.0) & \phantom{-1}48.9(8.0) & -0.160(7) & \phantom{-}0.028(1) & -0.163(7) & \phantom{-}0.029(1) \\
  & 3d\nicefrac{3}{2} & \phantom{1}-25.4(1.3) & \phantom{-1}25.4(1.3) & -0.161(6) & \phantom{-}0.028(1) & -0.163(6) & \phantom{-}0.029(1) \\
  & 3d\nicefrac{5}{2} & \phantom{-1}28.3(1.3) & \phantom{1}-28.3(1.3) & -0.103(3) & -0.027 & -0.103(3) & -0.027 \\[7pt]
  $^{147}$Sm & 1s\nicefrac{1}{2} & \phantom{-11}0 & \phantom{-11}0 & \text{\phantom{-}0.42(18)} & -0.33(14) & \phantom{-}0.25(17) & -0.20(14) \\
  & 2s\nicefrac{1}{2} & \phantom{-11}0 & \phantom{-11}0 & \phantom{-}0.072(39) & -0.056(30) & \phantom{-}0.033(39) & -0.026(30) \\
  & 2p\nicefrac{1}{2} & \phantom{-11}0 & \phantom{-11}0 & \phantom{-}0.164(58) & -0.127(45) & \phantom{-}0.106(58) & -0.082(45) \\
  & 2p\nicefrac{3}{2} & \phantom{1}-32.8(3.2) & \phantom{-1}32.8(3.2) & \phantom{-}0.066(8) & -0.004(1) & \phantom{-}0.058(8) & -0.004(1) \\
  & 3s\nicefrac{1}{2} & \phantom{-11}0 & \phantom{-11}0 & \phantom{-}0.023(13) & -0.018(10) & \phantom{-}0.010(13) & -0.008(8) \\
  & 3p\nicefrac{1}{2} & \phantom{-11}0 & \phantom{-11}0 & \phantom{-}0.044(18) & -0.034(14) & \phantom{-}0.026(18) & -0.02(1) \\
  & 3p\nicefrac{3}{2} & \phantom{11}-9.4(1.1) & \phantom{-11}9.4(1.1) & \phantom{-}0.020(3) & -0.001 & \phantom{-}0.017(3) & -0.001 \\
  & 3d\nicefrac{3}{2} & \phantom{11}-3.2(0.1) & \phantom{-11}3.2(0.1) & \phantom{-}0.015(1) & \phantom{-}0.000 & \phantom{-}0.014(1) & \phantom{-}0.000 \\
  & 3d\nicefrac{5}{2} & \phantom{-11}3.7(0.2) & \phantom{11}-3.7(0.2) & \phantom{-}0.010 & \phantom{-}0.004 & \phantom{-}0.010 & \phantom{-}0.004 \\[7pt]
 $^{89}$Zr & 1s\nicefrac{1}{2} & \phantom{-11}0 & \phantom{-11}0 & \phantom{-}0.36(13) & -0.29(10) & \phantom{-}0.23(12) & -0.19(10) \\
  & 2s\nicefrac{1}{2} & \phantom{-11}0 & \phantom{-11}0 & \phantom{-}0.053(23) & -0.043(18) & \phantom{-}0.030(23) & -0.025(18) \\
  & 2p\nicefrac{1}{2} & \phantom{-11}0 & \phantom{-11}0 & \phantom{-}0.071(14) & -0.058(11) & \phantom{-}0.057(14) & -0.047(11) \\
  & 2p\nicefrac{3}{2} & \phantom{-1}12.2(4.7) &\phantom{1}-12.2(4.7) & \phantom{-}0.023(1) & -0.004 & \phantom{-}0.022(1) &-0.004 \\
  & 3s\nicefrac{1}{2} & \phantom{-11}0 & \phantom{-11}0 & \phantom{-}0.016(7) & -0.013(6) & \phantom{-}0.009(7) & -0.007(6) \\
  & 3p\nicefrac{1}{2} & \phantom{-11}0 & \phantom{-11}0 & \phantom{-}0.020(4) & -0.017(4) & \phantom{-}0.016(4) & -0.013(4) \\
  & 3p\nicefrac{3}{2} & \phantom{-11}3.6(1.4) & \phantom{11}-3.6(1.4) & \phantom{-}0.007 & -0.001 & \phantom{-}0.007 & -0.001 \\
  & 3d\nicefrac{3}{2} & \text{\phantom{-11}0.9(0.3)} & \text{\phantom{11}-0.9(0.3)} & \phantom{-}0.004 & \phantom{-}0.000 & \phantom{-}0.004 & \phantom{-}0.000 \\
  & 3d\nicefrac{5}{2} & \text{\phantom{11}-1.1(0.4)} & \text{\phantom{-11}1.1(0.4)} & \phantom{-}0.003 & \phantom{-}0.000 & \phantom{-}0.003 &\phantom{-}0.000 \\

\end{tabular}
\end{ruledtabular}
\end{table*}
Since for heavy nuclei the nuclear radius is comparable to the muon's Compton wavelength \cite{Angeli2013,codata}, the muonic wavefunction overlaps strongly with the nucleus and the muon is sensitive to nuclear shape corrections, which results in hyperfine splitting of the energy levels. The quadrupole part of the electric interaction (\ref{eq:quadInt}) can be rewritten as \cite{Kozhedub2008}
\begin{equation}
\label{eq:Hquad}
H^{(2)}_E = - \alpha \frac{Q_0 F_{\text{QD}}(r_\mu)}{2\, r_\mu^3} \sum_{m=-2}^2 C_{2m}(\vartheta_N,\varphi_N)C_{2m}^{*}(\vartheta_\mu,\varphi_\mu),
\end{equation}
where $C_{lm}(\vartheta,\varphi)=\sqrt{4\pi/(2l+1)}Y_{lm}(\vartheta,\varphi)$ and angles with a subscript $\mu$($N$) describe the position of the muon ($z$ axis of the nuclear frame) in the laboratory frame. Here, the nuclear intrinsic quadrupole moment is defined via the charge distribution (\ref{eq:rho}) as
\begin{equation}
\label{eq:defQ0}
Q_0 = 2 \sqrt{\frac{4\pi}{5}} \int_0^\infty r^4 \rho_2(r)\,\mathrm{d}r,
\end{equation}
and the distribution of the quadrupole moment is described by the function $f(r_\mu)$, where in the point-like limit $f(r_\mu)=1/r_\mu^{-3}$. For the shell model, where the quadrupole distribution is concentrated around the nuclear rms radius $R_N$, the divergence for $r_\mu=0$ is removed, and the corresponding quadrupole distribution function is
\begin{equation}
F_{\text{QD}}(r_\mu)=
\begin{cases}
\left(\nicefrac{r_\mu}{R_N}\right)^5 & r_\mu \leq R_N\\
1 &r_\mu > R_N
\end{cases}.
\end{equation}
Formally, this corresponds to a charge distribution with
\begin{equation}
\rho_2(r_\mu)=\frac{Q_0}{2 R_N^4}\sqrt{\frac{5}{4\pi}}\delta(r_\mu-R_N).
\end{equation}
The matrix elements of the quadrupole interaction (\ref{eq:Hquad}) in the states (\ref{eq:totalState}) read \cite{Korzinin2005}
\begin{IEEEeqnarray}{l}
\label{eq:hquad}
\bra{FM_FI\kappa}H^{(2)}_E\ket{FM_FI\kappa}= - \alpha (-1)^{j+I+F}\\
\times \bra{I|}\frac{Q_0}{2} \widehat{C}_2(\vartheta_N,\varphi_N)\ket{|I} 
\bra{n\kappa|}\frac{F_{\text{QD}}(r_\mu)}{r_\mu^{3}}\widehat{C}_2(\vartheta_\mu,\varphi_\mu)\ket{|n\kappa}\nonumber.
\end{IEEEeqnarray}
The reduced matrix element in the nuclear coordinates can be expressed with the spectroscopic nuclear quadrupole moment $Q$ as
\begin{equation}
\nonumber
\bra{I|}\frac{Q_0}{2} \widehat{C}_2(\vartheta_N,\varphi_N)\ket{|I}= Q\sqrt{\frac{(2I+3)(2I+1)(I+1)}{4I(2I-1)}},
\end{equation}
and the reduced matrix elements in the muonic coordinates are 
\begin{IEEEeqnarray}{l}
\bra{n\kappa|}f(r_\mu)\widehat{C}_2(\vartheta_\mu,\varphi_\mu)\ket{|n\kappa} =\\
\quad-\sqrt{\frac{(2j+3)(2j+1)(2j-1)}{16j(j+1)}} \nonumber\\
\quad\times\int_0^\infty \left( G^2_{n\kappa}(r_\mu)+F^2_{n\kappa}(r_\mu)\right)\frac{F_{\text{QD}}(r_\mu)}{r_\mu^{3}} \mathrm{d}r_\mu.\nonumber
\end{IEEEeqnarray}
The values for the nuclear rms-radii $R_N$ and the spectroscopic quadrupole moments $Q$ are taken from Refs.~\cite{Angeli2013,Stone2005}. In Table \ref{tab:hfs}, results for the electric quadrupole hyperfine splitting for the nuclei $^{205}_{83}$Bi, $^{147}_{62}$Sm, and $^{89}_{40}$Zr are shown for a selection of hyperfine states, including uncertainties stemming from the error in the quadrupole moment and an estimation of the modeling uncertainty.
\subsection{Magnetic dipole splitting}
\label{sec:magndip}
In addition, the hyperfine splitting arises from the interaction of the nuclear magnetic moment with the muon's magnetic moment, which is also sensitive to the spatial distribution of the nuclear currents. Since the magnetic moment of the muon is inversely proportional to its mass, the magnetic hyperfine splitting in muonic atoms is less important than in electronic atoms. The matrix elements of the corresponding Hamiltonian (\ref{eq:Hmag}) in the state (\ref{eq:totalState}) are \cite{Korzinin2005}
\begin{IEEEeqnarray}{l}
\label{eq:hmag}
\bra{FM_FI\kappa}H_M\ket{FM_FI\kappa}=\\
\,\,\left[ F(F+1)-I(I+1)-j(j+1)\right] \nonumber\\
\,\,\times\frac{\alpha}{2 m_p}\frac{\mu}{\mu_N}\frac{\kappa}{Ij(j+1)}\int_0^\infty \frac{G_{n\kappa}(r_\mu)F_{n\kappa}(r_\mu)F_{\text{BW}}(r_\mu)}{r_\mu^2}\mathrm{d}r_\mu,\nonumber
\end{IEEEeqnarray}
where $m_p$ is the proton mass, and the ratio of the observed magnetic moment $\mu:=\bra{II}(\vec{\mu})_z\ket{II}$ and the nuclear magneton $\mu_N$ can be found in the literature \cite{Stone2005}. For the simple model of a homogeneous nuclear current distribution the distribution function (\ref{eq:Fbw}) of the Bohr-Weisskopf effect reads
\begin{equation}
\label{eq:bwsimple}
F_{\rm{BW}}(r_\mu)=\begin{cases}
\left( \nicefrac{r_\mu}{R_N} \right)^3 & r_\mu \leq R_N\\
1 &r_\mu > R_N
\end{cases}.
\end{equation}
Furthermore, an additional method is used to obtain the distribution function $F_{\rm{BW}}$ from the nuclear single particle model, where the nuclear magnetic moment is assigned to the odd nucleon and the Schrödinger equation for this nucleon is solved in the Woods-Saxon potential of the other nucleons \cite{Elizarov2005}. In Table \ref{tab:hfs}, results for the magnetic dipole hyperfine splitting for the nuclei $^{205}_{83}$Bi, $^{147}_{62}$Sm, and $^{89}_{40}$Zr are presented for a selection of hyperfine states, using both methods for obtaining $F_{\rm{BW}}$, where the model error is estimated by the difference of these two methods and the uncertainty in the magnetic moment is also taken into account.
\section{Conclusion}
Improved calculations for the fine and hyperfine structure of heavy muonic atoms were presented. In this work, finite-size corrections, leading-order vacuum polarization, electron screening, and nonrelativistic recoil corrections are already included in the solution of the Dirac equation. Thus, all further calculations of the hyperfine structure also contain these corrections via using the corrected wave functions. The electric quadrupole and magnetic dipole hyperfine structure was calculated to first order, using extended charge and current distributions. The detailed shape of these distributions represent a source of uncertainty for the predicted values, and thus motivates the comparison with experimental data, especially for nuclei with to date unknown charge distributions.

The presented usage of modified wave functions for the calculation of hyperfine effects can be extended to other phenomena in muonic atoms, for example, the dynamic hyperfine structure with highly deformed nuclei.
%

\begin{thebibliography}{31}%
\makeatletter
\providecommand \@ifxundefined [1]{%
 \@ifx{#1\undefined}
}%
\providecommand \@ifnum [1]{%
 \ifnum #1\expandafter \@firstoftwo
 \else \expandafter \@secondoftwo
 \fi
}%
\providecommand \@ifx [1]{%
 \ifx #1\expandafter \@firstoftwo
 \else \expandafter \@secondoftwo
 \fi
}%
\providecommand \natexlab [1]{#1}%
\providecommand \enquote  [1]{``#1''}%
\providecommand \bibnamefont  [1]{#1}%
\providecommand \bibfnamefont [1]{#1}%
\providecommand \citenamefont [1]{#1}%
\providecommand \href@noop [0]{\@secondoftwo}%
\providecommand \href [0]{\begingroup \@sanitize@url \@href}%
\providecommand \@href[1]{\@@startlink{#1}\@@href}%
\providecommand \@@href[1]{\endgroup#1\@@endlink}%
\providecommand \@sanitize@url [0]{\catcode `\\12\catcode `\$12\catcode
  `\&12\catcode `\#12\catcode `\^12\catcode `\_12\catcode `\%12\relax}%
\providecommand \@@startlink[1]{}%
\providecommand \@@endlink[0]{}%
\providecommand \url  [0]{\begingroup\@sanitize@url \@url }%
\providecommand \@url [1]{\endgroup\@href {#1}{\urlprefix }}%
\providecommand \urlprefix  [0]{URL }%
\providecommand \Eprint [0]{\href }%
\providecommand \doibase [0]{http://dx.doi.org/}%
\providecommand \selectlanguage [0]{\@gobble}%
\providecommand \bibinfo  [0]{\@secondoftwo}%
\providecommand \bibfield  [0]{\@secondoftwo}%
\providecommand \translation [1]{[#1]}%
\providecommand \BibitemOpen [0]{}%
\providecommand \bibitemStop [0]{}%
\providecommand \bibitemNoStop [0]{.\EOS\space}%
\providecommand \EOS [0]{\spacefactor3000\relax}%
\providecommand \BibitemShut  [1]{\csname bibitem#1\endcsname}%
\let\auto@bib@innerbib\@empty
\bibitem [{\citenamefont {Mohr}\ \emph {et~al.}(2016)\citenamefont {Mohr},
  \citenamefont {Newell},\ and\ \citenamefont {Taylor}}]{codata}%
  \BibitemOpen
  \bibfield  {author} {\bibinfo {author} {\bibfnamefont {P.~J.}\ \bibnamefont
  {Mohr}}, \bibinfo {author} {\bibfnamefont {D.~B.}\ \bibnamefont {Newell}}, \
  and\ \bibinfo {author} {\bibfnamefont {B.~N.}\ \bibnamefont {Taylor}},\
  }\href {\doibase 10.1103/RevModPhys.88.035009} {\bibfield  {journal}
  {\bibinfo  {journal} {Rev. Mod. Phys.}\ }\textbf {\bibinfo {volume} {88}},\
  \bibinfo {pages} {035009} (\bibinfo {year} {2016})}\BibitemShut {NoStop}%
\bibitem [{\citenamefont {Wheeler}(1949)}]{Wheeler1949}%
  \BibitemOpen
  \bibfield  {author} {\bibinfo {author} {\bibfnamefont {J.~A.}\ \bibnamefont
  {Wheeler}},\ }\href {\doibase 10.1103/RevModPhys.21.133} {\bibfield
  {journal} {\bibinfo  {journal} {Rev. Mod. Phys.}\ }\textbf {\bibinfo {volume}
  {21}},\ \bibinfo {pages} {133} (\bibinfo {year} {1949})}\BibitemShut
  {NoStop}%
\bibitem [{\citenamefont {Borie}\ and\ \citenamefont
  {Rinker}(1982)}]{BorieRinker1982}%
  \BibitemOpen
  \bibfield  {author} {\bibinfo {author} {\bibfnamefont {E.}~\bibnamefont
  {Borie}}\ and\ \bibinfo {author} {\bibfnamefont {G.~A.}\ \bibnamefont
  {Rinker}},\ }\href {\doibase 10.1103/RevModPhys.54.67} {\bibfield  {journal}
  {\bibinfo  {journal} {Rev. Mod. Phys.}\ }\textbf {\bibinfo {volume} {54}},\
  \bibinfo {pages} {67} (\bibinfo {year} {1982})}\BibitemShut {NoStop}%
\bibitem [{\citenamefont {Devons}\ and\ \citenamefont
  {Duerdoth}(1995)}]{Devons1995}%
  \BibitemOpen
  \bibfield  {author} {\bibinfo {author} {\bibfnamefont {S.}~\bibnamefont
  {Devons}}\ and\ \bibinfo {author} {\bibfnamefont {I.}~\bibnamefont
  {Duerdoth}},\ }\enquote {\bibinfo {title} {Muonic atoms},}\ in\ \href
  {\doibase 10.1007/978-1-4684-8343-7_5} {\emph {\bibinfo {booktitle} {Advances
  in Nuclear Physics: Volume 2}}},\ \bibinfo {editor} {edited by\ \bibinfo
  {editor} {\bibfnamefont {M.}~\bibnamefont {Baranger}}\ and\ \bibinfo {editor}
  {\bibfnamefont {E.}~\bibnamefont {Vogt}}}\ (\bibinfo  {publisher} {Springer
  US},\ \bibinfo {address} {Boston, MA},\ \bibinfo {year} {1995})\ pp.\
  \bibinfo {pages} {295--423}\BibitemShut {NoStop}%
\bibitem [{\citenamefont {Piller}\ \emph {et~al.}(1990)\citenamefont {Piller},
  \citenamefont {Gugler}, \citenamefont {Jacot-Guillarmod}, \citenamefont
  {Schaller}, \citenamefont {Schellenberg}, \citenamefont {Schneuwly},
  \citenamefont {Fricke}, \citenamefont {Hennemann},\ and\ \citenamefont
  {Herberz}}]{Piller1990}%
  \BibitemOpen
  \bibfield  {author} {\bibinfo {author} {\bibfnamefont {C.}~\bibnamefont
  {Piller}}, \bibinfo {author} {\bibfnamefont {C.}~\bibnamefont {Gugler}},
  \bibinfo {author} {\bibfnamefont {R.}~\bibnamefont {Jacot-Guillarmod}},
  \bibinfo {author} {\bibfnamefont {L.~A.}\ \bibnamefont {Schaller}}, \bibinfo
  {author} {\bibfnamefont {L.}~\bibnamefont {Schellenberg}}, \bibinfo {author}
  {\bibfnamefont {H.}~\bibnamefont {Schneuwly}}, \bibinfo {author}
  {\bibfnamefont {G.}~\bibnamefont {Fricke}}, \bibinfo {author} {\bibfnamefont
  {T.}~\bibnamefont {Hennemann}}, \ and\ \bibinfo {author} {\bibfnamefont
  {J.}~\bibnamefont {Herberz}},\ }\href {\doibase 10.1103/PhysRevC.42.182}
  {\bibfield  {journal} {\bibinfo  {journal} {Phys. Rev. C}\ }\textbf {\bibinfo
  {volume} {42}},\ \bibinfo {pages} {182} (\bibinfo {year} {1990})}\BibitemShut
  {NoStop}%
\bibitem [{\citenamefont {Schaller}\ \emph {et~al.}(1980)\citenamefont
  {Schaller}, \citenamefont {Schellenberg}, \citenamefont {Ruetschi},\ and\
  \citenamefont {Schneuwly}}]{Schaller1980}%
  \BibitemOpen
  \bibfield  {author} {\bibinfo {author} {\bibfnamefont {L.}~\bibnamefont
  {Schaller}}, \bibinfo {author} {\bibfnamefont {L.}~\bibnamefont
  {Schellenberg}}, \bibinfo {author} {\bibfnamefont {A.}~\bibnamefont
  {Ruetschi}}, \ and\ \bibinfo {author} {\bibfnamefont {H.}~\bibnamefont
  {Schneuwly}},\ }\href {\doibase
  http://dx.doi.org/10.1016/0375-9474(80)90656-9} {\bibfield  {journal}
  {\bibinfo  {journal} {Nuclear Physics A}\ }\textbf {\bibinfo {volume}
  {343}},\ \bibinfo {pages} {333 } (\bibinfo {year} {1980})}\BibitemShut
  {NoStop}%
\bibitem [{\citenamefont {Dey}\ \emph {et~al.}(1979)\citenamefont {Dey},
  \citenamefont {Ebersold}, \citenamefont {Leisi}, \citenamefont {Scheck},
  \citenamefont {Walter},\ and\ \citenamefont {Zehnder}}]{Dey1979}%
  \BibitemOpen
  \bibfield  {author} {\bibinfo {author} {\bibfnamefont {W.}~\bibnamefont
  {Dey}}, \bibinfo {author} {\bibfnamefont {P.}~\bibnamefont {Ebersold}},
  \bibinfo {author} {\bibfnamefont {H.}~\bibnamefont {Leisi}}, \bibinfo
  {author} {\bibfnamefont {F.}~\bibnamefont {Scheck}}, \bibinfo {author}
  {\bibfnamefont {H.}~\bibnamefont {Walter}}, \ and\ \bibinfo {author}
  {\bibfnamefont {A.}~\bibnamefont {Zehnder}},\ }\href {\doibase
  http://dx.doi.org/10.1016/0375-9474(79)90401-9} {\bibfield  {journal}
  {\bibinfo  {journal} {Nuclear Physics A}\ }\textbf {\bibinfo {volume}
  {326}},\ \bibinfo {pages} {418 } (\bibinfo {year} {1979})}\BibitemShut
  {NoStop}%
\bibitem [{\citenamefont {Rüetschi}\ \emph {et~al.}(1984)\citenamefont
  {Rüetschi}, \citenamefont {Schellenberg}, \citenamefont {Phan},
  \citenamefont {Piller}, \citenamefont {Schaller},\ and\ \citenamefont
  {Schneuwly}}]{Ruetschi1984}%
  \BibitemOpen
  \bibfield  {author} {\bibinfo {author} {\bibfnamefont {A.}~\bibnamefont
  {Rüetschi}}, \bibinfo {author} {\bibfnamefont {L.}~\bibnamefont
  {Schellenberg}}, \bibinfo {author} {\bibfnamefont {T.}~\bibnamefont {Phan}},
  \bibinfo {author} {\bibfnamefont {G.}~\bibnamefont {Piller}}, \bibinfo
  {author} {\bibfnamefont {L.}~\bibnamefont {Schaller}}, \ and\ \bibinfo
  {author} {\bibfnamefont {H.}~\bibnamefont {Schneuwly}},\ }\href {\doibase
  http://dx.doi.org/10.1016/0375-9474(84)90359-2} {\bibfield  {journal}
  {\bibinfo  {journal} {Nuclear Physics A}\ }\textbf {\bibinfo {volume}
  {422}},\ \bibinfo {pages} {461 } (\bibinfo {year} {1984})}\BibitemShut
  {NoStop}%
\bibitem [{\citenamefont {Bergem}\ \emph {et~al.}(1988)\citenamefont {Bergem},
  \citenamefont {Piller}, \citenamefont {Rueetschi}, \citenamefont {Schaller},
  \citenamefont {Schellenberg},\ and\ \citenamefont {Schneuwly}}]{Bergem1988}%
  \BibitemOpen
  \bibfield  {author} {\bibinfo {author} {\bibfnamefont {P.}~\bibnamefont
  {Bergem}}, \bibinfo {author} {\bibfnamefont {G.}~\bibnamefont {Piller}},
  \bibinfo {author} {\bibfnamefont {A.}~\bibnamefont {Rueetschi}}, \bibinfo
  {author} {\bibfnamefont {L.~A.}\ \bibnamefont {Schaller}}, \bibinfo {author}
  {\bibfnamefont {L.}~\bibnamefont {Schellenberg}}, \ and\ \bibinfo {author}
  {\bibfnamefont {H.}~\bibnamefont {Schneuwly}},\ }\href {\doibase
  10.1103/PhysRevC.37.2821} {\bibfield  {journal} {\bibinfo  {journal} {Phys.
  Rev. C}\ }\textbf {\bibinfo {volume} {37}},\ \bibinfo {pages} {2821}
  (\bibinfo {year} {1988})}\BibitemShut {NoStop}%
\bibitem [{\citenamefont {Pohl~\textit{et al.}}(2010)}]{Pohl2010}%
  \BibitemOpen
  \bibfield  {author} {\bibinfo {author} {\bibfnamefont {R.}~\bibnamefont
  {Pohl~\textit{et al.}}},\ }\href {\doibase 10.1038/nature09250} {\bibfield
  {journal} {\bibinfo  {journal} {Nature}\ }\textbf {\bibinfo {volume} {466}},\
  \bibinfo {pages} {213} (\bibinfo {year} {2010})}\BibitemShut {NoStop}%
\bibitem [{\citenamefont {Indelicato}(2013)}]{indelicato2013}%
  \BibitemOpen
  \bibfield  {author} {\bibinfo {author} {\bibfnamefont {P.}~\bibnamefont
  {Indelicato}},\ }\href {\doibase 10.1103/PhysRevA.87.022501} {\bibfield
  {journal} {\bibinfo  {journal} {Phys. Rev. A}\ }\textbf {\bibinfo {volume}
  {87}},\ \bibinfo {pages} {022501} (\bibinfo {year} {2013})}\BibitemShut
  {NoStop}%
\bibitem [{\citenamefont {Pachucki}\ and\ \citenamefont
  {Wienczek}(2015)}]{pachucki2015}%
  \BibitemOpen
  \bibfield  {author} {\bibinfo {author} {\bibfnamefont {K.}~\bibnamefont
  {Pachucki}}\ and\ \bibinfo {author} {\bibfnamefont {A.}~\bibnamefont
  {Wienczek}},\ }\href {\doibase 10.1103/PhysRevA.91.040503} {\bibfield
  {journal} {\bibinfo  {journal} {Phys. Rev. A}\ }\textbf {\bibinfo {volume}
  {91}},\ \bibinfo {pages} {040503} (\bibinfo {year} {2015})}\BibitemShut
  {NoStop}%
\bibitem [{\citenamefont {Wansbeek}\ \emph {et~al.}(2008)\citenamefont
  {Wansbeek}, \citenamefont {Sahoo}, \citenamefont {Timmermans}, \citenamefont
  {Jungmann}, \citenamefont {Das},\ and\ \citenamefont
  {Mukherjee}}]{Wansbeek2008}%
  \BibitemOpen
  \bibfield  {author} {\bibinfo {author} {\bibfnamefont {L.~W.}\ \bibnamefont
  {Wansbeek}}, \bibinfo {author} {\bibfnamefont {B.~K.}\ \bibnamefont {Sahoo}},
  \bibinfo {author} {\bibfnamefont {R.~G.~E.}\ \bibnamefont {Timmermans}},
  \bibinfo {author} {\bibfnamefont {K.}~\bibnamefont {Jungmann}}, \bibinfo
  {author} {\bibfnamefont {B.~P.}\ \bibnamefont {Das}}, \ and\ \bibinfo
  {author} {\bibfnamefont {D.}~\bibnamefont {Mukherjee}},\ }\href {\doibase
  10.1103/PhysRevA.78.050501} {\bibfield  {journal} {\bibinfo  {journal} {Phys.
  Rev. A}\ }\textbf {\bibinfo {volume} {78}},\ \bibinfo {pages} {050501}
  (\bibinfo {year} {2008})}\BibitemShut {NoStop}%
\bibitem [{\citenamefont {{Kirch}}(2016)}]{kirch2016}%
  \BibitemOpen
  \bibfield  {author} {\bibinfo {author} {\bibfnamefont {K.}~\bibnamefont
  {{Kirch}}},\ }\href@noop {} {\bibfield  {journal} {\bibinfo  {journal} {ArXiv
  e-prints}\ } (\bibinfo {year} {2016})},\ \Eprint
  {http://arxiv.org/abs/1607.07042} {arXiv:1607.07042 [hep-ph]} \BibitemShut
  {NoStop}%
\bibitem [{\citenamefont {Shabaev}\ \emph {et~al.}(2004)\citenamefont
  {Shabaev}, \citenamefont {Tupitsyn}, \citenamefont {Yerokhin}, \citenamefont
  {Plunien},\ and\ \citenamefont {Soff}}]{Shabaev2004}%
  \BibitemOpen
  \bibfield  {author} {\bibinfo {author} {\bibfnamefont {V.~M.}\ \bibnamefont
  {Shabaev}}, \bibinfo {author} {\bibfnamefont {I.~I.}\ \bibnamefont
  {Tupitsyn}}, \bibinfo {author} {\bibfnamefont {V.~A.}\ \bibnamefont
  {Yerokhin}}, \bibinfo {author} {\bibfnamefont {G.}~\bibnamefont {Plunien}}, \
  and\ \bibinfo {author} {\bibfnamefont {G.}~\bibnamefont {Soff}},\ }\href
  {\doibase 10.1103/PhysRevLett.93.130405} {\bibfield  {journal} {\bibinfo
  {journal} {Phys. Rev. Lett.}\ }\textbf {\bibinfo {volume} {93}},\ \bibinfo
  {pages} {130405} (\bibinfo {year} {2004})}\BibitemShut {NoStop}%
\bibitem [{\citenamefont {Kozhedub}\ \emph {et~al.}(2008)\citenamefont
  {Kozhedub}, \citenamefont {Andreev}, \citenamefont {Shabaev}, \citenamefont
  {Tupitsyn}, \citenamefont {Brandau}, \citenamefont {Kozhuharov},
  \citenamefont {Plunien},\ and\ \citenamefont {St\"ohlker}}]{Kozhedub2008}%
  \BibitemOpen
  \bibfield  {author} {\bibinfo {author} {\bibfnamefont {Y.~S.}\ \bibnamefont
  {Kozhedub}}, \bibinfo {author} {\bibfnamefont {O.~V.}\ \bibnamefont
  {Andreev}}, \bibinfo {author} {\bibfnamefont {V.~M.}\ \bibnamefont
  {Shabaev}}, \bibinfo {author} {\bibfnamefont {I.~I.}\ \bibnamefont
  {Tupitsyn}}, \bibinfo {author} {\bibfnamefont {C.}~\bibnamefont {Brandau}},
  \bibinfo {author} {\bibfnamefont {C.}~\bibnamefont {Kozhuharov}}, \bibinfo
  {author} {\bibfnamefont {G.}~\bibnamefont {Plunien}}, \ and\ \bibinfo
  {author} {\bibfnamefont {T.}~\bibnamefont {St\"ohlker}},\ }\href {\doibase
  10.1103/PhysRevA.77.032501} {\bibfield  {journal} {\bibinfo  {journal} {Phys.
  Rev. A}\ }\textbf {\bibinfo {volume} {77}},\ \bibinfo {pages} {032501}
  (\bibinfo {year} {2008})}\BibitemShut {NoStop}%
\bibitem [{\citenamefont {Hitlin}\ \emph {et~al.}(1970)\citenamefont {Hitlin},
  \citenamefont {Bernow}, \citenamefont {Devons}, \citenamefont {Duerdoth},
  \citenamefont {Kast}, \citenamefont {Macagno}, \citenamefont {Rainwater},
  \citenamefont {Wu},\ and\ \citenamefont {Barrett}}]{hitlin1970}%
  \BibitemOpen
  \bibfield  {author} {\bibinfo {author} {\bibfnamefont {D.}~\bibnamefont
  {Hitlin}}, \bibinfo {author} {\bibfnamefont {S.}~\bibnamefont {Bernow}},
  \bibinfo {author} {\bibfnamefont {S.}~\bibnamefont {Devons}}, \bibinfo
  {author} {\bibfnamefont {I.}~\bibnamefont {Duerdoth}}, \bibinfo {author}
  {\bibfnamefont {J.~W.}\ \bibnamefont {Kast}}, \bibinfo {author}
  {\bibfnamefont {E.~R.}\ \bibnamefont {Macagno}}, \bibinfo {author}
  {\bibfnamefont {J.}~\bibnamefont {Rainwater}}, \bibinfo {author}
  {\bibfnamefont {C.~S.}\ \bibnamefont {Wu}}, \ and\ \bibinfo {author}
  {\bibfnamefont {R.~C.}\ \bibnamefont {Barrett}},\ }\href {\doibase
  10.1103/PhysRevC.1.1184} {\bibfield  {journal} {\bibinfo  {journal} {Phys.
  Rev. C}\ }\textbf {\bibinfo {volume} {1}},\ \bibinfo {pages} {1184} (\bibinfo
  {year} {1970})}\BibitemShut {NoStop}%
\bibitem [{\citenamefont {Johnson}(2007)}]{johnson2007}%
  \BibitemOpen
  \bibfield  {author} {\bibinfo {author} {\bibfnamefont {W.~R.}\ \bibnamefont
  {Johnson}},\ }\href@noop {} {\emph {\bibinfo {title} {Atomic Structure
  Theory}}},\ \bibinfo {edition} {1st}\ ed.\ (\bibinfo  {publisher}
  {Springer},\ \bibinfo {address} {Berlin Heidelberg},\ \bibinfo {year}
  {2007})\BibitemShut {NoStop}%
\bibitem [{\citenamefont {Elizarov}\ \emph {et~al.}(2005)\citenamefont
  {Elizarov}, \citenamefont {Shabaev}, \citenamefont {Oreshkina},\ and\
  \citenamefont {Tupitsyn}}]{Elizarov2005}%
  \BibitemOpen
  \bibfield  {author} {\bibinfo {author} {\bibfnamefont {A.~A.}\ \bibnamefont
  {Elizarov}}, \bibinfo {author} {\bibfnamefont {V.~M.}\ \bibnamefont
  {Shabaev}}, \bibinfo {author} {\bibfnamefont {N.~S.}\ \bibnamefont
  {Oreshkina}}, \ and\ \bibinfo {author} {\bibfnamefont {I.~I.}\ \bibnamefont
  {Tupitsyn}},\ }\href
  {http://www.sciencedirect.com/science/article/pii/S0168583X0500371X}
  {\bibfield  {journal} {\bibinfo  {journal} {Nucl. Instrum. Methods Phys. Res.
  B}\ }\textbf {\bibinfo {volume} {235}},\ \bibinfo {pages} {65} (\bibinfo
  {year} {2005})}\BibitemShut {NoStop}%
\bibitem [{\citenamefont {Bohr}\ and\ \citenamefont
  {Weisskopf}(1950)}]{bohrWeisskopf1950}%
  \BibitemOpen
  \bibfield  {author} {\bibinfo {author} {\bibfnamefont {A.}~\bibnamefont
  {Bohr}}\ and\ \bibinfo {author} {\bibfnamefont {V.~F.}\ \bibnamefont
  {Weisskopf}},\ }\href {\doibase 10.1103/PhysRev.77.94} {\bibfield  {journal}
  {\bibinfo  {journal} {Phys. Rev.}\ }\textbf {\bibinfo {volume} {77}},\
  \bibinfo {pages} {94} (\bibinfo {year} {1950})}\BibitemShut {NoStop}%
\bibitem [{\citenamefont {Steffen}(1985)}]{Steffen1985}%
  \BibitemOpen
  \bibfield  {author} {\bibinfo {author} {\bibfnamefont {R.~M.}\ \bibnamefont
  {Steffen}},\ }\href@noop {} {\bibfield  {journal} {\bibinfo  {journal}
  {Hyperfine Interactions}\ }\textbf {\bibinfo {volume} {24}},\ \bibinfo
  {pages} {223} (\bibinfo {year} {1985})}\BibitemShut {NoStop}%
\bibitem [{\citenamefont {Varshalovich}\ \emph {et~al.}(1988)\citenamefont
  {Varshalovich}, \citenamefont {Moskalev},\ and\ \citenamefont
  {Khersonskii}}]{varshalovich1988}%
  \BibitemOpen
  \bibfield  {author} {\bibinfo {author} {\bibfnamefont {D.~A.}\ \bibnamefont
  {Varshalovich}}, \bibinfo {author} {\bibfnamefont {A.~N.}\ \bibnamefont
  {Moskalev}}, \ and\ \bibinfo {author} {\bibfnamefont {V.~K.}\ \bibnamefont
  {Khersonskii}},\ }\href@noop {} {\emph {\bibinfo {title} {Quantum Theory of
  Angular Momentum}}}\ (\bibinfo  {publisher} {World Scientific},\ \bibinfo
  {address} {Singapore},\ \bibinfo {year} {1988})\BibitemShut {NoStop}%
\bibitem [{\citenamefont {Beier}(2000)}]{Beier2000}%
  \BibitemOpen
  \bibfield  {author} {\bibinfo {author} {\bibfnamefont {T.}~\bibnamefont
  {Beier}},\ }\href {\doibase http://dx.doi.org/10.1016/S0370-1573(00)00071-5}
  {\bibfield  {journal} {\bibinfo  {journal} {Physics Reports}\ }\textbf
  {\bibinfo {volume} {339}},\ \bibinfo {pages} {79 } (\bibinfo {year}
  {2000})}\BibitemShut {NoStop}%
\bibitem [{\citenamefont {Angeli}\ and\ \citenamefont
  {Marinova}(2013)}]{Angeli2013}%
  \BibitemOpen
  \bibfield  {author} {\bibinfo {author} {\bibfnamefont {I.}~\bibnamefont
  {Angeli}}\ and\ \bibinfo {author} {\bibfnamefont {K.}~\bibnamefont
  {Marinova}},\ }\href {\doibase http://dx.doi.org/10.1016/j.adt.2011.12.006}
  {\bibfield  {journal} {\bibinfo  {journal} {Atomic Data and Nuclear Data
  Tables}\ }\textbf {\bibinfo {volume} {99}},\ \bibinfo {pages} {69 } (\bibinfo
  {year} {2013})}\BibitemShut {NoStop}%
\bibitem [{\citenamefont {Greiner}(2000)}]{greiner2000}%
  \BibitemOpen
  \bibfield  {author} {\bibinfo {author} {\bibfnamefont {W.}~\bibnamefont
  {Greiner}},\ }\href@noop {} {\emph {\bibinfo {title} {Relativistic Quantum
  Mechanics}}},\ \bibinfo {edition} {3rd}\ ed.\ (\bibinfo  {publisher}
  {Springer-Verlag},\ \bibinfo {address} {Berlin Heidelberg},\ \bibinfo {year}
  {2000})\BibitemShut {NoStop}%
\bibitem [{\citenamefont {Landau}\ and\ \citenamefont
  {Lifshitz}(1981)}]{landaulifshitz3}%
  \BibitemOpen
  \bibfield  {author} {\bibinfo {author} {\bibfnamefont {L.~D.}\ \bibnamefont
  {Landau}}\ and\ \bibinfo {author} {\bibfnamefont {L.~M.}\ \bibnamefont
  {Lifshitz}},\ }\href {http://www.worldcat.org/isbn/0750635398} {\emph
  {\bibinfo {title} {Quantum Mechanics Non-Relativistic Theory, Third Edition:
  Volume 3}}},\ \bibinfo {edition} {3rd}\ ed.\ (\bibinfo  {publisher}
  {Butterworth-Heinemann},\ \bibinfo {year} {1981})\BibitemShut {NoStop}%
\bibitem [{\citenamefont {Friar}\ and\ \citenamefont
  {Negele}(1973)}]{friar1973}%
  \BibitemOpen
  \bibfield  {author} {\bibinfo {author} {\bibfnamefont {J.}~\bibnamefont
  {Friar}}\ and\ \bibinfo {author} {\bibfnamefont {J.}~\bibnamefont {Negele}},\
  }\href {\doibase http://dx.doi.org/10.1016/0370-2693(73)90459-0} {\bibfield
  {journal} {\bibinfo  {journal} {Physics Letters B}\ }\textbf {\bibinfo
  {volume} {46}},\ \bibinfo {pages} {5 } (\bibinfo {year} {1973})}\BibitemShut
  {NoStop}%
\bibitem [{\citenamefont {Vogel}(1973)}]{vogel1973}%
  \BibitemOpen
  \bibfield  {author} {\bibinfo {author} {\bibfnamefont {P.}~\bibnamefont
  {Vogel}},\ }\href {\doibase 10.1103/PhysRevA.7.63} {\bibfield  {journal}
  {\bibinfo  {journal} {Phys. Rev. A}\ }\textbf {\bibinfo {volume} {7}},\
  \bibinfo {pages} {63} (\bibinfo {year} {1973})}\BibitemShut {NoStop}%
\bibitem [{\citenamefont {Bethe}\ and\ \citenamefont
  {Salpeter}(1977)}]{bethe_salpeter}%
  \BibitemOpen
  \bibfield  {author} {\bibinfo {author} {\bibfnamefont {H.}~\bibnamefont
  {Bethe}}\ and\ \bibinfo {author} {\bibfnamefont {E.}~\bibnamefont
  {Salpeter}},\ }\href@noop {} {\emph {\bibinfo {title} {Quantum Mechanics of
  One- and Two-Electron Systems}}}\ (\bibinfo  {publisher} {Plenum Publishing
  Corporation},\ \bibinfo {year} {1977})\BibitemShut {NoStop}%
\bibitem [{\citenamefont {Korzinin}\ \emph {et~al.}(2005)\citenamefont
  {Korzinin}, \citenamefont {Oreshkina},\ and\ \citenamefont
  {Shabaev}}]{Korzinin2005}%
  \BibitemOpen
  \bibfield  {author} {\bibinfo {author} {\bibfnamefont {E.~Y.}\ \bibnamefont
  {Korzinin}}, \bibinfo {author} {\bibfnamefont {N.~S.}\ \bibnamefont
  {Oreshkina}}, \ and\ \bibinfo {author} {\bibfnamefont {V.~M.}\ \bibnamefont
  {Shabaev}},\ }\href {http://stacks.iop.org/1402-4896/71/i=5/a=008} {\bibfield
   {journal} {\bibinfo  {journal} {Physica Scripta}\ }\textbf {\bibinfo
  {volume} {71}},\ \bibinfo {pages} {464} (\bibinfo {year} {2005})}\BibitemShut
  {NoStop}%
\bibitem [{\citenamefont {Stone}(2005)}]{Stone2005}%
  \BibitemOpen
  \bibfield  {author} {\bibinfo {author} {\bibfnamefont {N.~J.}\ \bibnamefont
  {Stone}},\ }\href {\doibase http://dx.doi.org/10.1016/j.adt.2005.04.001}
  {\bibfield  {journal} {\bibinfo  {journal} {Atomic Data and Nuclear Data
  Tables}\ }\textbf {\bibinfo {volume} {90}},\ \bibinfo {pages} {75 } (\bibinfo
  {year} {2005})}\BibitemShut {NoStop}%
\end{thebibliography}
%
%
%
\end{document}